\documentclass{PoS}
\usepackage{float}
\usepackage{subfig}
\usepackage{graphicx}

\title{Exploring the Phase Diagram of Lattice Quantum Gravity}

\ShortTitle{Exploring the Phase Diagram of Lattice Quantum Gravity}

\author{\speaker{Daniel Coumbe}\\
  SUPA, School of Physics and Astronomy, University of Glasgow, Glasgow, G12 8QQ, UK\\
  E-mail: \email{d.coumbe@physics.gla.ac.uk}}

\author{Jack Laiho\\  
  SUPA, School of Physics and Astronomy, University of Glasgow, Glasgow, G12 8QQ, UK\\
  E-mail: \email{j.laiho@physics.gla.ac.uk}}

\abstract{We present evidence that a nonperturbative model of quantum gravity defined via Euclidean dynamical triangulations contains a region in parameter space with an extended 4-dimensional geometry when a non-trivial measure term is included in the gravitational path integral. Within our extended region we find a large scale spectral dimension of \begin{math}D_{s}\left(\sigma\rightarrow\infty\right)=4.04\pm0.26\end{math} and a Hausdorff dimension that is consistent with \begin{math}D_{H}=4\end{math} from finite size scaling. We find that the short distance spectral dimension is \begin{math}D_{s}\left(\sigma\rightarrow0\right)\approx3/2\end{math}, which may resolve the tension between asymptotic safety and holographic entropy scaling.} 

\FullConference{XXIX International Symposium on Lattice Field Theory \\
                July 10-16 2011\\
                Squaw Valley, Lake Tahoe, California}

\begin{document}

\section{Introduction}
The quantization of gravity is one of the great outstanding problems in theoretical physics. A straightforward implementation of general relativity as a perturbative quantum field theory is not renormalizable by power counting. Although it is still possible to formulate gravity as an effective field theory at low energies \cite{Donoghue97}, at each order in the perturbative expansion, new divergences appear, requiring an infinite number of couplings that encode the physics at high energy scales, leading to a loss of predictive power. However, the conjecture that gravity could be nonperturbatively renormalizable was made by Weinberg in Ref. \cite{Weinberg76}, where it was termed the asymptotic safety scenario. In this scenario, the renormalization group flow of couplings end at a non-trivial fixed point in the ultraviolet, with a finite dimensional critical surface. Gravity, therefore, would be safe from ultraviolet divergences and describable in terms of a finite number of parameters, making it effectively renormalizable when formulated nonperturbatively.

A lattice formulation allows one to define a gravitational path integral that can be studied nonperturbatively. One of the original formulations of lattice gravity is Euclidean dynamical triangulations (EDT) \cite{Ambjorn92,Catterall94}, which defines a spacetime of locally flat n-simplices of fixed edge length, where an n-simplex is the n-dimensional analogue of a triangle. An ensemble can be generated using the Metropolis algorithm with a set of local update moves. The partition function we use is  


\begin{equation}
Z_{E}={\sum_{T}}\frac{1}{C_{T}}\left[\prod_{j=1}^{N_{2}}\mathcal{O}\left(t_{j}\right)^{\beta}\right]e^{-S_{E}},
\end{equation}

\noindent where the product is over all 2-simplices, and \begin{math}\mathcal{O}\left(t_{j}\right)\end{math} is the order of the 2-simplex \emph{j}, i.e. the number of four-simplices to which the triangle belongs. The term in square brackets is a non-trivial measure term. We vary the free parameter \begin{math}\beta\end{math} as an additional independent coupling constant in the bare lattice action \cite{Brugmann92}.
The Einstein-Regge action is given by,


\begin{equation}
S_{E}=-\kappa_{2}N_{2}+\kappa_{4}N_{4},
\end{equation}

\noindent where \begin{math}\kappa_{2}\end{math} and \begin{math}\kappa_{4}\end{math} are related to the bare Newton's constant and the cosmological constant, respectively. \begin{math}\kappa_{4}\end{math} must be tuned to its critical value such that an infinite volume limit can be taken \cite{deBakker94}. This leaves a 2-dimensional parameter space that can be explored by varying \begin{math}\kappa_{2}\end{math} and \begin{math}\beta\end{math}.

Early studies of EDT typically included only two couplings in the bare lattice action, \begin{math}\kappa_{2}\end{math} and \begin{math}\kappa_{4}\end{math}, and they typically did not allow for a non-trivial measure term. The original EDT model ran into significant problems. The parameter space of couplings contained just two phases, neither of which resembled 4-dimensional semi-classical general relativity, and the two phases were seperated by a first order critical point, making it unlikely that one could take a continuum limit. In response to these problems a causality condition was added by Ambjorn and Loll \cite{AmbjornLoll98}. The causality condition distinguishes between space-like and time-like links on the lattice so that an explicit foliation of the lattice into space-like hypersufaces can be introduced, where all hypersurfaces have the same topology. Only geometries that can be foliated in this way are included in the ensemble of triangulations that define the path integral measure. A 4-dimensional de Sitter like phase was shown to emerge from causal dynamical triangulations (CDT) \cite{AmbjornLoll98}\cite{AmbjornGorlich08}. 

In this work we revisit the original Euclidean theory with a non-trivial measure term for two reasons. One is that the CDT restriction to a fixed foliation is potentially at odds with general covariance, though it may amount to a choice of gauge. We would like to see whether the successes of the CDT formulation can be reproduced using the explicitly covariant EDT formulation. Second, renormalization group studies \cite{Codello09,Benedetti09} suggest that the ultraviolet critical surface is 3-dimensional, providing the original motivation for revisiting EDT with a third parameter in the bare lattice action \cite{Brugmann92,Bilke98}. We point out that the CDT action also includes a third parameter, an asymmetry parameter between space and time. The results from our EDT approach obtained so far suggest that the causality constraint of CDT is not necessary to obtain a good four-dimensional semiclassical limit \cite{Laiho11}. 

\section{Phase Diagram}

The new parameter \begin{math}\beta\end{math} associated with the non-trivial measure term enlarges the parameter space in which to explore the phase diagram. The three different regions of the phase diagram are labelled collapsed, branched polymer and extended.  


  \begin{figure}[h]
    \centering
    \includegraphics[width=0.4\linewidth]{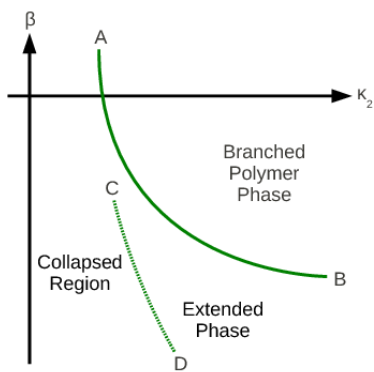}
    \caption{\small A schematic of the phase diagram as a function of $\kappa_{2}$ and $\beta$.}
  \end{figure}

\noindent The collapsed region is characterised by a fractal Hausdorff dimension \begin{math}D_{H}\rightarrow\infty\end{math}. The branched polymer phase has a Hausdorff dimension \begin{math}D_{H}=2\end{math} and a spectral dimension \begin{math}D_{s}=4/3\end{math}. These two regions are unphysical. However, there exists a region in the parameter space with an extended 4-dimensional geometry. The solid line AB dividing the extended and branched polymer phases has been extensively studied in the literature at the point $\beta=0$. At that point the transition is almost certaintly first order \cite{deBakker96,Bialas96}. The dashed line CD appears to be much softer than the AB transition and appears to be a cross-over seperating the collapsed and extended regions of a single phase.   

\section{Numerical Implementation and Code Tests}

A d-dimensional simplicial manifold is constructed by gluing d-simplices together along their (d-1)-dimensional faces. To each d-simplex there exists a simplex label and a set of combinatorially unique (d+1) vertex labels. The set of combinatorial triangulations was used in most early simulations of EDT. However, the constraint of combinatorial uniqueness can be relaxed to include the larger set of degenerate triangulations in which the neighbours of a given simplex are no longer unique \cite{Bilke99}. It has been shown numerically that simulations using degenerate triangulations lead to a factor of \begin{math}\sim10\end{math} reduction in finite size effects compared to combinatorial triangulations \cite{Bilke99}. We have made various checks of our code against the literature using combinatorial triangulations \cite{deBakker95}, as well as for degenerate triangulations \cite{Bilke98}, and good agreement was found. 

\section{Exploring the Phase Diagram}

Determining the critical behaviour of the phase diagram is of great importance because the existence of a second order critical point allows one to define a continuum limit for the theory. The Regge curvature \begin{math} R \end{math} serves as an order parameter for exploring the phase diagram, with a derivative with respect to \begin{math}\kappa_{2}\end{math} given by the susceptability \begin{math}\chi_{R}\left(N_{2},N_{4}\right)\end{math}. 


\begin{equation}
\left\langle R \right\rangle=\frac{1}{\rho} \left(\frac{N_{2}}{N_{4}}\right) - 1,
\end{equation}


\begin{equation}
\chi_{R}(N_{2},N_{4})=\left[\left\langle \left(\frac{N_{2}}{N_{4}}\right)^{2}\right\rangle -\left\langle \frac{N_{2}}{N_{4}}\right\rangle ^{2}\right]N_{4},
\end{equation}

\noindent where \begin{math}\rho=\frac{10\arccos\left(1/4\right)}{2\pi}\end{math} and \begin{math}N_{i}\end{math} is the total number of \emph{i}-simplices. We have studied the phase transition line AB and the transition region CD by calculating the average Regge curvature and curvature susceptability for fixed values of \begin{math}\beta\end{math}, whilst varying \begin{math}\kappa_{2}\end{math}. 


  \begin{figure}[h]
    \centering
    \subfloat[]{\includegraphics[width=0.5\textwidth]{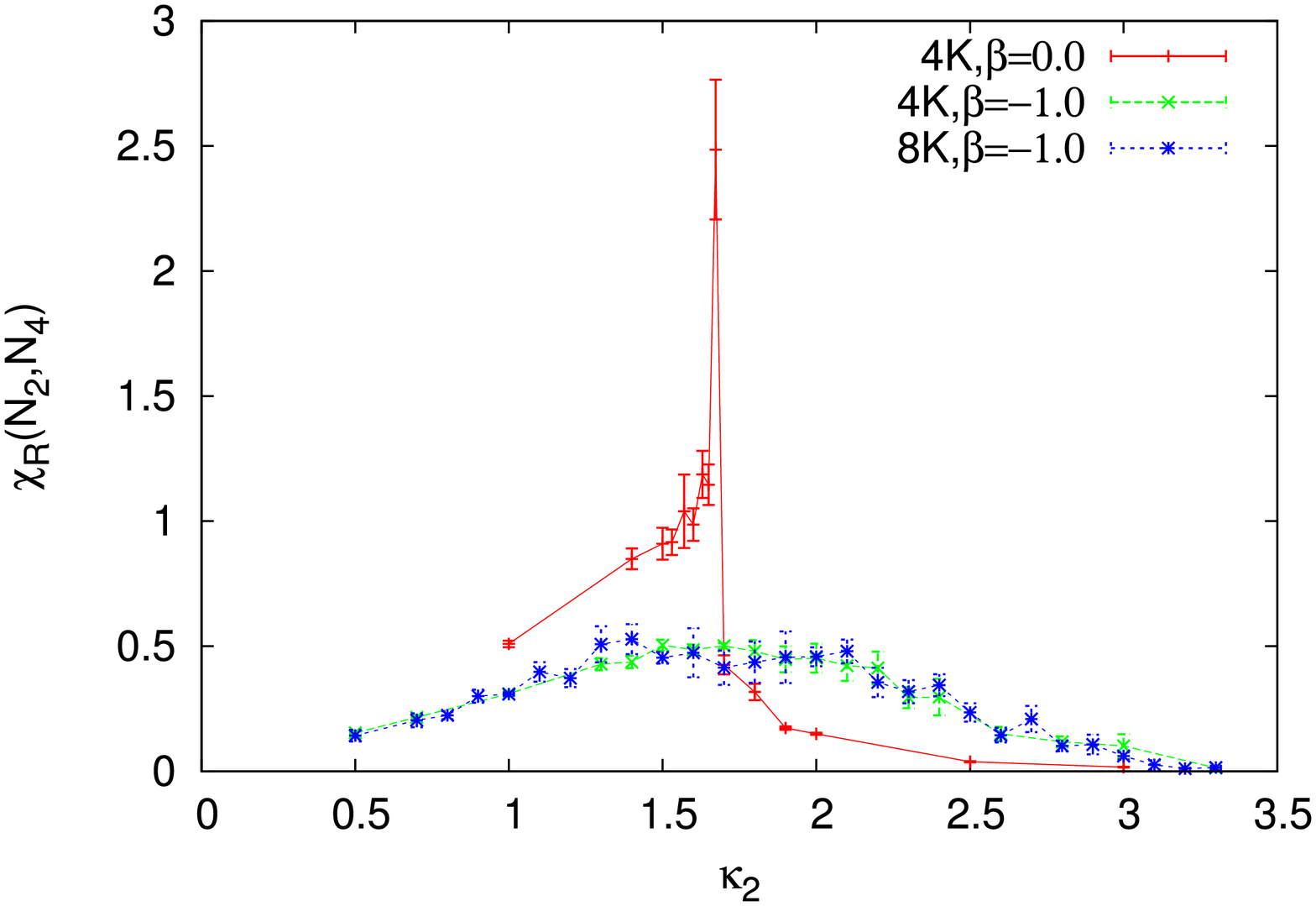}\label{fig:chir}}\subfloat[]{\includegraphics[width=0.5\textwidth]{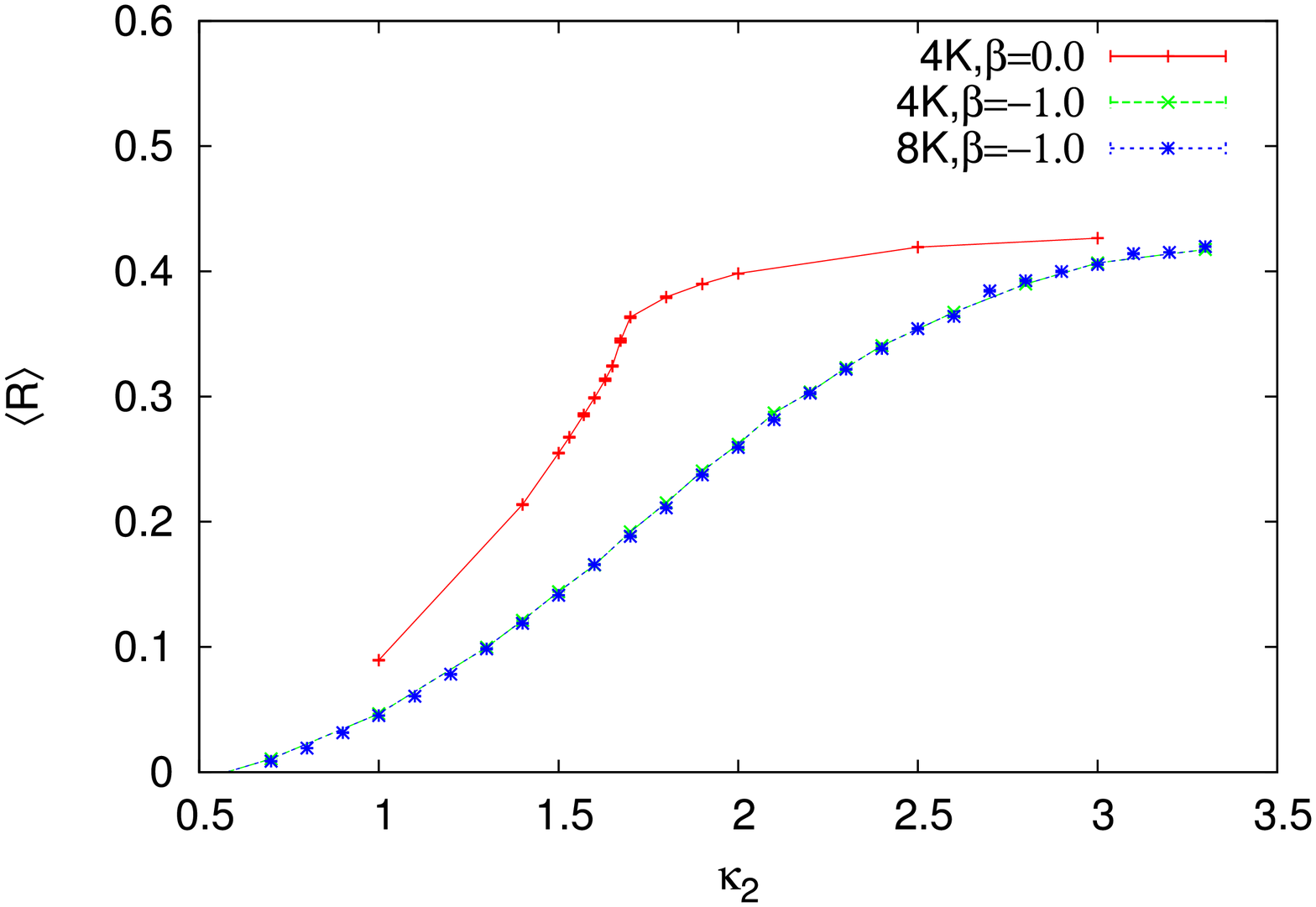}\label{fig:avgr3}}             
    \caption{\small Simulation using 4000 degenerate simplices for $\beta=0$ and 4000 and 8000 for $\beta=-1.0$, whilst varying $\kappa_{2}$. (a) The curvature susceptability as a function of $\kappa_{2}$. (b) The average Regge curvature as a function of $\kappa_{2}$.}
  \end{figure}

Figure \ref{fig:chir} shows a sharp peak in the curvature susceptability for \begin{math}\beta = 0\end{math}, and a less sharp peak for \begin{math}\beta=-1.0\end{math}. The behaviour at \begin{math}\beta= -1.0\end{math} is consistent with a cross-over as the peak in the curvature susceptability does not appear to change with volume.\interfootnotelinepenalty=10000 \footnote{\scriptsize The transition could be third or higher order, since it is very difficult to distinguish an analytic cross-over from a higher order transition using numerical methods.}    

It is well established in the literature \cite{deBakker96} that, at the point $\beta=0$, along the critical line AB, there is a first order phase transition for combinatorial triangulations, and Ref. \cite{Thorleifsson98} provides evidence that this is also true for degenerate triangulations. We show the Monte Carlo time history of \begin{math}N_{0}\end{math} and the corresponding histogram for the ensemble at the peak in \begin{math}\chi_{R}\end{math} at \begin{math}\kappa_{2}=1.672\end{math} and \begin{math}\beta=0\end{math}.


\begin{figure}[h]
  \centering
  \subfloat[]{\includegraphics[width=0.49\textwidth]{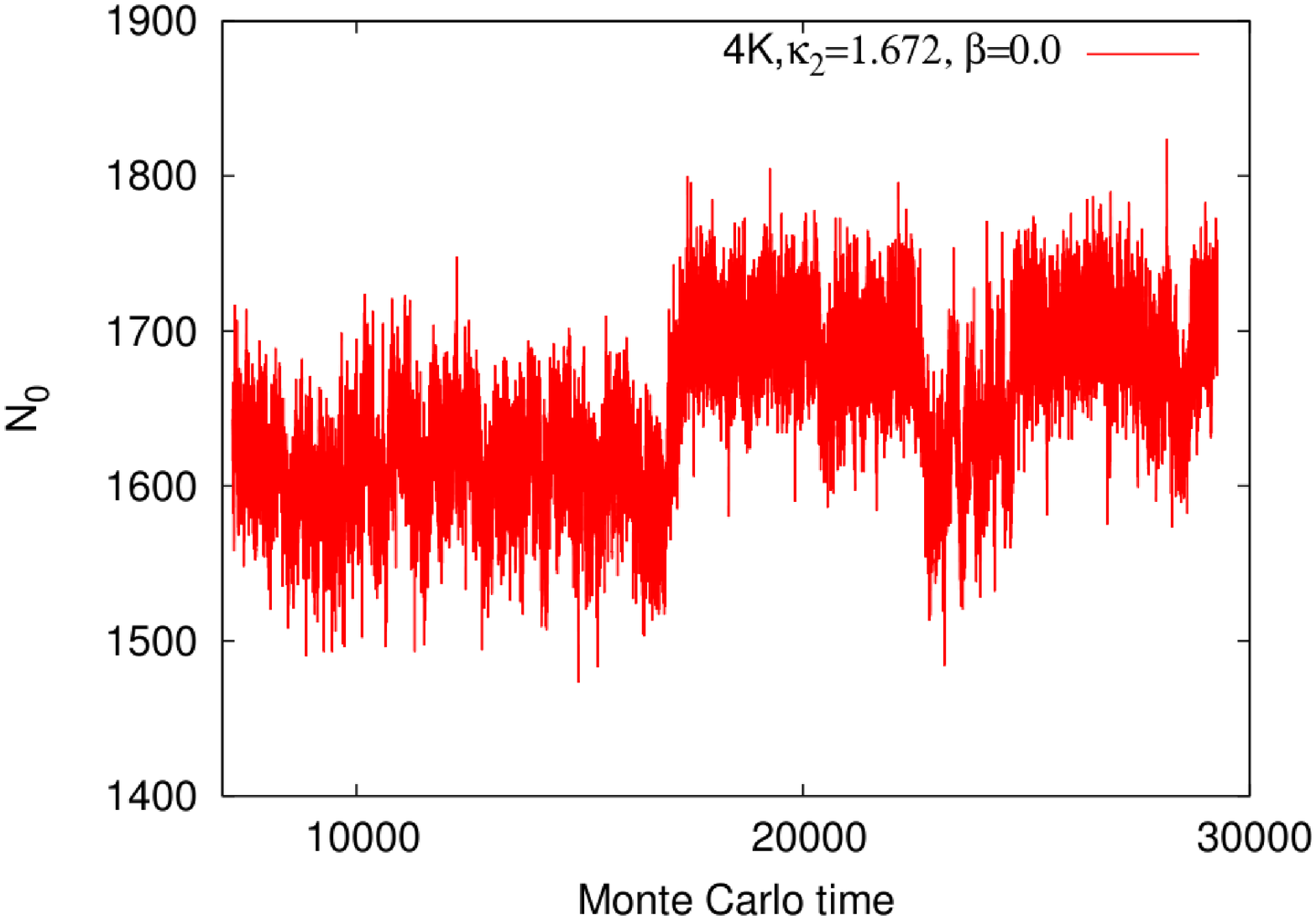}\label{fig:mctime}}\subfloat[]{\includegraphics[width=0.5\textwidth]{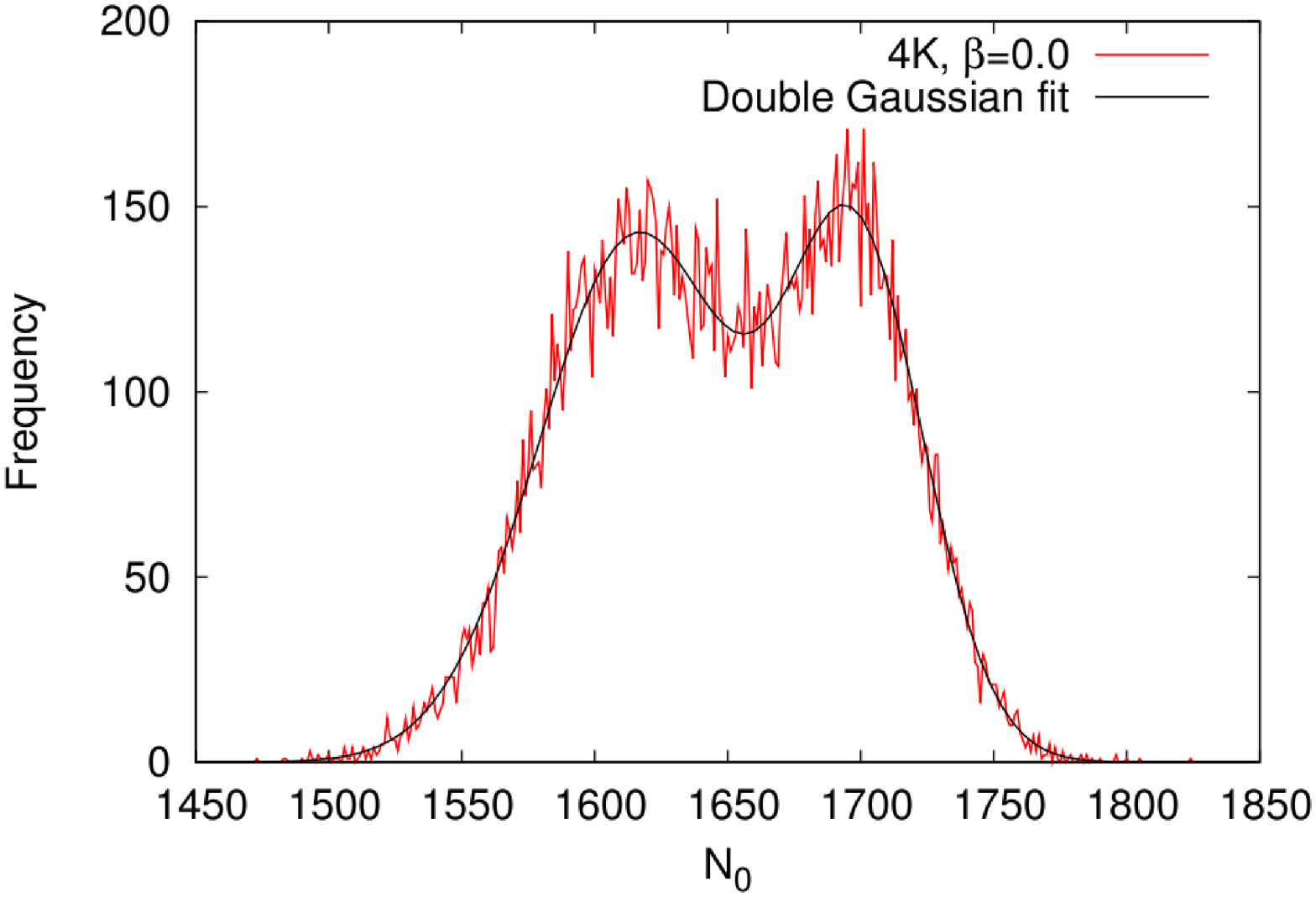}\label{fig:histogram}}             
  \caption{\small Simulation using 4000 degenerate simplices at $\kappa_{2}=1.672$ and $\beta=0$. (a) The number of vertices, $N_{0}$, as a function of Monte Carlo time. (b) Histogram of $N_{0}$ values. }
\end{figure}
     
\noindent Figure \ref{fig:mctime} shows evidence for discontinuous fluctuations between the two metastable values of \begin{math}N_{0}\end{math} that are characteristic of a discontinuous phase transition, and Figure \ref{fig:histogram}  shows the histogram of \begin{math}N_{0}\end{math}. A first order phase transition has a signature double Gaussian peak in the histogram of \begin{math}N_{0}\end{math}, as seen in Figure \ref{fig:histogram}.       

\section{The Extended Region}

\begin{figure}[h]
    \centering
    \subfloat[]{\includegraphics[width=0.45\textwidth]{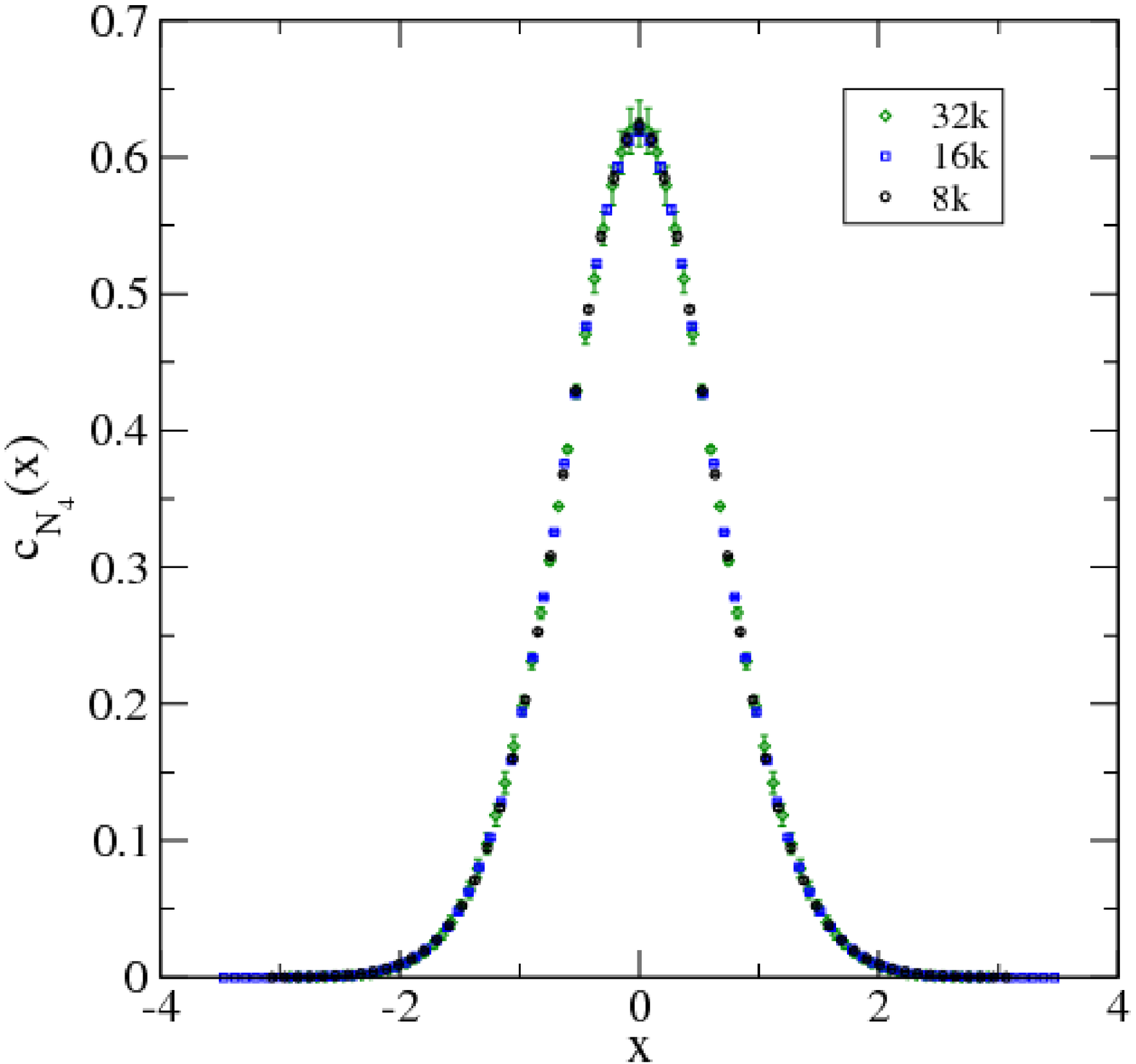}\label{fig:volcorr}}\subfloat[]{\includegraphics[width=0.439\textwidth]{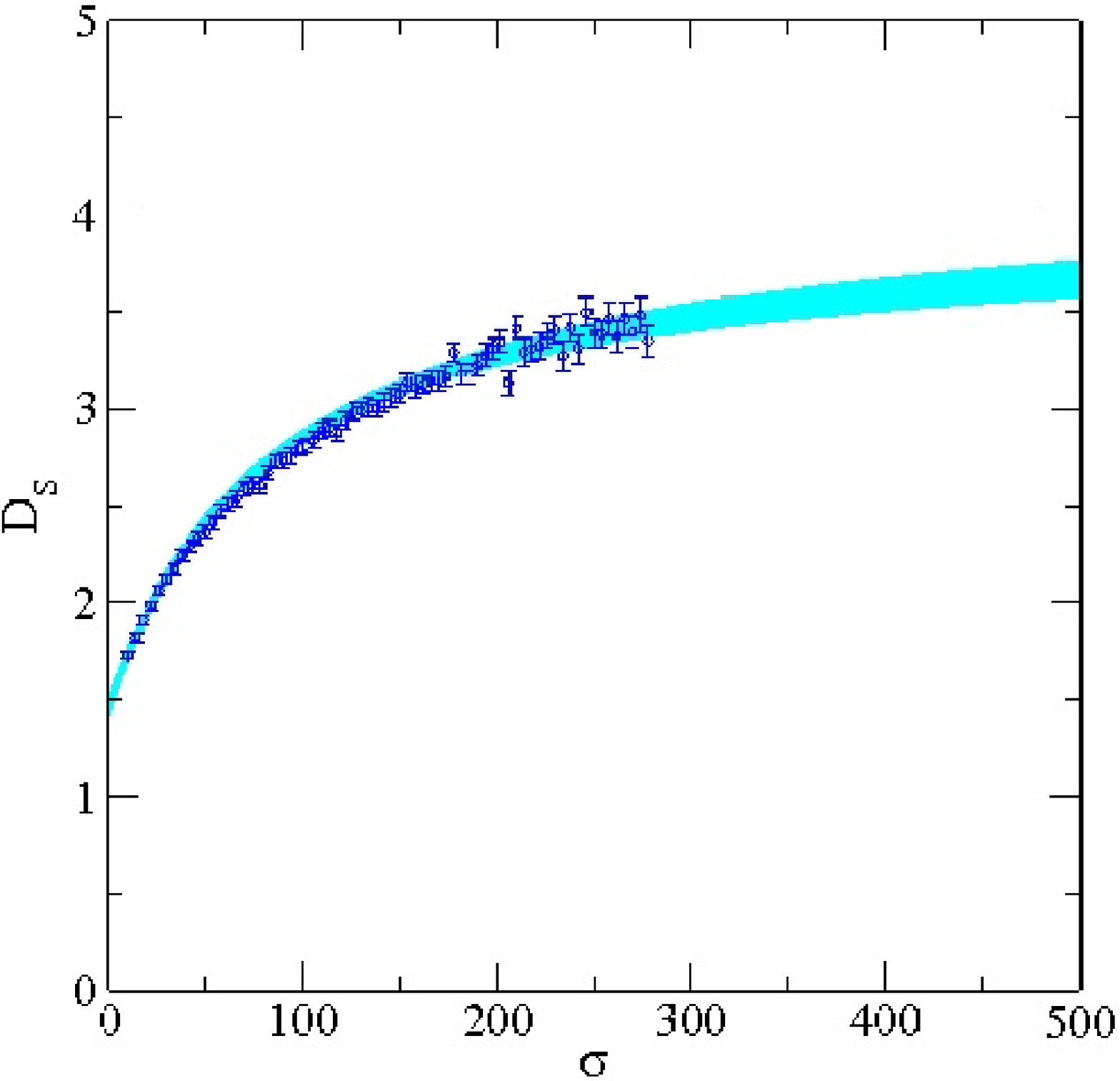}\label{fig:spec}}             
    \caption{\small (a) Three-volume correlator using 8000, 16000 and 32000 four simplices. (b) The running spectral dimension as a function of diffusion time, including a fit function of the form $D_{s}=a-\frac{b}{c+\sigma}$.}
  \end{figure}

Simulations investigating finite-size scaling support the fact that there exists a region in parameter space where our EDT model has a 4-dimensional extended geometry. We study the finite volume scaling behaviour of the three-volume correlator similar to the one introduced in Ref. \cite{Ambjorn05} to study the scaling of CDT, 


\begin{equation}
C_{N_{4}}\left(\delta\right)=\sum_{\tau=1}^t\frac{\left\langle N_{4}^{\rm slice}(\tau)N_{4}^{\rm slice}(\tau+\delta)\right\rangle }{N_{4}^{2}}.
\end{equation}

\noindent \begin{math}N_{4}^{\rm slice}(\tau)\end{math} is the total number of 4-simplices in a spherical shell a geodesic distance \begin{math}\tau\end{math} from a randomly chosen simplex. \begin{math}N_{4}\end{math} is the total number of 4-simplices and the normalization of the correlator is chosen such that \begin{math}\sum_{\delta=0}^{t-1}C_{N_{4}}\left(\delta\right)=1.\end{math} If we rescale \begin{math}\delta\end{math} and \begin{math}C_{N_{4}}\left(\delta\right)\end{math}, defining \begin{math}x=\delta/N_{4}^{1/D_{H}}\end{math}, then the universal distribution \begin{math}c_{N_{4}}(x)\end{math} should be independent of the lattice volume, where \begin{math}c_{N_{4}}\left(x\right)=N_{4}^{1/D_{H}}C_{N_{4}}\left(\delta/N_{4}^{1/D_{H}}\right)\end{math}. One can determine the fractal Hausdorff dimension, \begin{math}D_{H}\end{math}, as the value that leaves \begin{math}c_{N_{4}}\left(x\right)\end{math} invariant under a change in four-volume \begin{math}N_{4}\end{math}. Good agreement between \begin{math}c_{N_{4}}\left(x\right)\end{math} at different volumes \begin{math}N_{4}\end{math} occurs when \begin{math}D_{H}=4\end{math}; thus the geometries have Hausdorff dimension close to 4. This is illustrated in Figure \ref{fig:volcorr}. 


The spectral dimension, \begin{math}D_{s}\end{math}, is related to the probability of return, \begin{math}P_{r}(\sigma)\end{math}, for a random walk over the ensemble of triangulations after \begin{math}\sigma\end{math} diffusion steps. It is defined by,


\begin{equation}
D_{s}=-2\frac{d\log P_{r}\left(\sigma\right)}{d\log\sigma}.
\end{equation}

\noindent Figure \ref{fig:spec} shows a plot of \begin{math}D_{s}\end{math} as a function of \begin{math}\sigma\end{math}, including a fit to the function \begin{math}D_{s}=a-\frac{b}{c+\sigma}\end{math} suggested in Ref. \cite{Ambjorn05b}. Our preferred fit to the spectral dimension gives \begin{math}D_{s}\left(\sigma\rightarrow\infty\right)=4.04\pm0.26\end{math} and \begin{math}D_{s}\left(\sigma\rightarrow0\right)=1.457\pm0.064\end{math}, where the error includes the statistical error and a systematic error associated with varying the fitting range and the fit function added to the statistical error in quadrature. Variations in the fit function assume that the spectral dimension approaches a constant in the \begin{math}\sigma\rightarrow\infty\end{math} limit, and that \begin{math}D_{s}\left(\sigma\right)\end{math} is monotonic. Although this result for the spectral dimension contains an estimate of the systematic error, to bring this error fully under control we will need to explore finite size effects for \begin{math}D_{s}\left(\sigma\rightarrow\infty\right)\end{math} and discretization effects for \begin{math}D_{s}\left(\sigma\rightarrow0\right)\end{math}. Work in progress using ensembles with larger volumes and longer diffusion times still give \begin{math}D_{s}\left(\sigma\rightarrow\infty\right)\end{math} consistent with four. It is especially important to quantify a systematic error associated with discretization effects, because our result for \begin{math}D_{s}\left(\sigma\rightarrow0\right)\end{math} is inconsistent with the renormalization group result \begin{math}D_{s}\left(\sigma\rightarrow0\right)=2\end{math} exactly \cite{Lauscher05}.\interfootnotelinepenalty=10000 \footnote{\scriptsize Ref. \protect\cite{Reuter11} argues using renormalization group methods that there will be a long plateau at \begin{math}\sim4/3\end{math} in \begin{math}D_{s}\left(\sigma\right)\end{math} for small values of \begin{math}\sigma\end{math} before \begin{math}D_{s}\end{math} increases to 2 as \begin{math}\sigma\rightarrow0\end{math}. Thus, they propose that our value \begin{math}D_{s}\left(\sigma\rightarrow0\right)\sim 3/2\end{math} may be the result of our using insufficiently fine lattice spacings and a non-monotonic behaviour of \begin{math}D_s\left(\sigma\right)\end{math} for small \begin{math}\sigma\end{math}.} So far, calculations at finer lattice spacings extrapolate to values close to \begin{math}D_{s}\left(\sigma\rightarrow0\right)=3/2\end{math} and show no sign of increasing or decreasing.\interfootnotelinepenalty=10000 \footnote{\scriptsize The relative lattice spacing is determined from a comparison of the running spectral dimension at different values of the bare parameters.} 

\section{Conclusions}

We have presented evidence that EDT with a non-trivial measure term has a region with 4-dimensional geometry. This geometry has a large scale spectral dimension of approximately four, and a Hausdorff dimension that is consistent with \begin{math}D_{H}=4,\end{math} as determined from finite size scaling. These results suggest that the explicitly covariant EDT approach to quantum gravity may be in the same universality class as CDT. The small scale spectral dimension of \begin{math}D_{s}\left(\sigma\rightarrow0\right)=1.457\pm0.064\end{math} is consistent with 3/2, and the CDT result \begin{math}D_{s}\left(\sigma\rightarrow0\right)=1.80\pm0.25\end{math} \cite{Ambjorn05} is also consistent with 3/2, given the current error estimate. As discussed in our earlier paper \cite{Laiho11}, this result may resolve the tension between asymptotic safety and holographic entropy scaling. 
   
\emph{This work was funded by STFC and the Scottish Universities Physics Alliance. Computing was done on Scotgrid and the DiRAC facility jointly funded by STFC and BIS.} 


\end{document}